\documentclass[fleqn,10pt]{wlscirep}
\usepackage[utf8]{inputenc}
\usepackage[T1]{fontenc}
\usepackage{bm}

\title{Anomalies in Particle Physics}

\author[1,2,*]{Andreas Crivellin}
\affil[1]{Physik-Institut, Universität Z\"urich, Winterthurerstrasse 190, CH–8057 Zürich, Switzerland}
\affil[2]{Paul Scherrer Institut, CH–5232 Villigen PSI, Switzerland}

\affil[*]{e-mail: andreas.crivellin@psi.ch}

\author[3,4,**]{Bruce Mellado}
\affil[3]{School of Physics and Institute for Collider Particle Physics, University of the Witwatersrand,
Johannesburg, Wits 2050, South Africa}
\affil[4]{iThemba LABS, National Research Foundation, PO Box 722, Somerset West 7129, South Africa}
\affil[**]{e-mail:bmellado@mail.cern.ch}

\begin{abstract}
The currently accepted mathematical description of the fundamental constituents and interactions of matter is the Standard Model of particle physics. Its last missing particle, the famous Higgs boson, was observed at the Large Hadron Collider at CERN in 2012. However, it is clear that the Standard Model cannot be the ultimate theory of Nature, and, e.g., cannot account for Dark Matter or non-vanishing neutrino masses (and does not include gravity). In fact, searches for physics beyond the SM have been intensified since the Higgs boson discovery. In this article, we review the hints for new physics, called ``anomalies'', obtained in particle physics experiments within the last years. We consider both direct high-energy searches for new resonances at the LHC and indirect low-energy precision experiments. These anomalies range from the nuclear scale (approximately the mass of the proton) to the electroweak scale (i.e.~the mass of the Higgs boson) to the TeV scale (the highest scale directly accessible at the LHC), therefore spanning over four orders of magnitude. After discussing the experimental and theoretical status of the anomalies, we summarize possible explanations in terms of new particles and new interactions. In particular, new Higgs bosons and leptoquarks are promising candidates. Discovery prospects and implications for future colliders are discussed.
\end{abstract}
\begin{document}

\flushbottom
\maketitle

\thispagestyle{empty}

\noindent \textbf{Key points:} 

\begin{itemize}
    \item The Standard Model of Particle Physics is the currently accepted mathematical theory describing the fundamental constituents of matter as well as their interactions, and was completed by the discovery of the Higgs particle at the Large Hadron Collider at CERN in 2012.
    \item The Standard Model cannot account for the existence of Dark Matter or non-vanishing neutrino masses and must therefore be extended, however, a plethora of viable options exist.
    \item In recent years, several interesting deviations from the Standard Model predictions, called ``anomalies'', were found, both in high-energy searches at the LHC and in low-energy precision observables.
    \item These anomalies range from precision measurements of properties of the muon, to hints for new scalar bosons at the electroweak scale, to the existence of heavy TeV scale resonances.
    \item {The anomalies} can be explained by supplementing the Standard Model with new particles and new interactions, in particular, additional Higgs bosons, new fermions and new strongly interacting particles.
    \item While the data of the third run of the Large Hadron Collider will already be able to establish the existence of {some of these} new particles if one or more of the anomalies are actually due to new physics, the high luminosity LHC and future colliders, like FCC-ee, ILC, CEPC or CLIC, as well as new precision experiments, are needed for a comprehensive and precise study of their properties. 
\end{itemize}

\noindent \textbf{Website summary:} 

The Standard Model of particle physics is the currently accepted theory of the fundamental constituents of matter and their interactions. We review the status of experimental hints for new physics, which, if confirmed, would require the extension of the Standard Model by new particles and new interactions.

\section{Introduction}

The Standard Model (SM) of particle physics is the currently accepted mathematical description of the fundamental constituents of matter and their interactions (excluding gravity). More specifically, matter consists of quarks and leptons which are fermions (spin 1/2 particles whose wave functions are invariant under {spatial} rotation of $4\pi$). A proton contains two up-quarks ($u$ with electric charge $+2/3$) and one down-quark ($d$ with charge $-1/3$), while a neutron consists of one up-quark and two down-quarks. Electrons ($e$) constitute the atomic shell. Together with the nearly mass-less and very weakly interacting neutrinos ($\nu$), they form the class called leptons. All fermions appear in three copies, called generations or flavours, that only differ in mass.\footnote{{To be more precise, all differences between the different generations of fermions are induced by the mass terms originating from their couplings to the Higgs particle. However, in the mass eigenbasis, this leads, in addition to the different mass eigenvalues, to flavour-specific couplings of the $W$ boson to quarks.}} The electron is accompanied by its heavy cousins the muon ($\mu$) and the even heavier tau ($\tau$). The more massive versions of the up-quark are called charm ($c$) and top ($t$) while strange ($s$) and bottom quark ($b$ sometimes also called ``beauty'') are the heavier copies of the down-quark. The masses of the charged fermions range from $\approx0.0005\,$GeV for the electron to $\approx 174\,$GeV for the top.\footnote{Note that a giga-electronvolt (GeV) is approximately the mass of a proton and thus of a hydrogen atom. While the up and down quark masses are very small compared to the proton mass, most of it is due to the binding energy of the strong interactions.} Only first-generation fermions are stable, while the heavier generations are short-lived and decay to lighter flavours. 

The forces between the fermions are mediated by {``gauge interactions''. They are based on `local'' symmetries, meaning that they hold independently at any point in space-time. The corresponding gauge group of the SM is $SU(3)_c\times SU(2)_L\times U(1)_Y$, which corresponds to rotations in 3, 2 and 1-dimensional (external) complex spaces. Due to quantization, i.e.~the wave/particle duality of quantum mechanics, these interactions result in force particles, the gauge bosons with spin one. The electromagnetic force is mediated by the photon ($\gamma$), the weak force  (corresponding to the $SU(2)_L$ factor) by the charged $W$ and the neutral $Z$ gauge bosons\footnote{{To be more precise, the photon and the $Z$ boson are linear combinations of the $U(1)_Y$ boson and the neutral component of $SU(2)_L$ gauge field.}} and the strong force ($SU(3)_c$) by eight gluons ($g$). {While neutrinos only feel the weak force, the charged leptons (electron, muon and tau) also have electromagnetic interactions and quarks are charged in addition under the strong force and thus interact with gluons.} Importantly, all flavour violation in the SM is induced by the couplings of the $W$ boson to up-type and down-type quarks via the Cabibbo-Kobayashi-Maskawa (CKM) matrix which accounts for the corresponding coupling strength.

Finally, we have the famous Higgs particle~\cite{Higgs:1964ia,Englert:1964et} ($h$), which was discovered at the Large Hadron Collider at CERN in 2012~\cite{ATLAS:2012yve,CMS:2012qbp}. It is the first and only (so far) fundamental scalar particle (spin 0) and the field from which the Higgs boson originates spontaneously breaks $SU(2)_L\times U(1)_Y$ to the electromagnetic gauge group ($U(1)_{\rm EM}$) with the mass-less photon. At the same time, it gives masses to the $W$  ($\approx80\,$GeV) and $Z$ bosons ($\approx90\,$GeV) as well as to all (fundamental) fermions and the Higgs boson itself ($\approx125\,$GeV). The full particle content of the SM is summarized in Fig.~\ref{StandardModel}.

\begin{figure}
	\centering
	\vspace{-3mm}
   \includegraphics[width=0.5\textwidth]{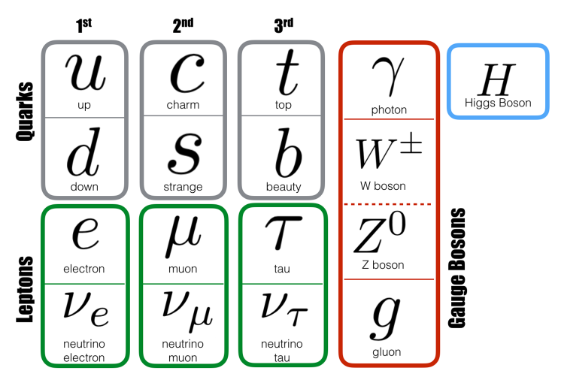}
	\vspace{-3mm}
	\caption{Particle content of the Standard Model: Fermions consisting of quarks (grey) and leptons (green). The forces are mediated by the gauge bosons (red) and the Higgs particle (blue) gives masses to the $W$ and $Z$ bosons as well as to all quarks and leptons.}
	\label{StandardModel}
\end{figure}

While the SM is now complete, it cannot be the ultimate fundamental theory of Nature: In addition to many theoretical arguments for the existence of beyond the SM (BSM) physics, the SM e.g.~cannot account for the observations of Dark Matter (DM) established at cosmological scales (since it does not contain a weakly interacting particle with the right relic abundance), nor for the non-vanishing neutrino masses\footnote{{Here we define the SM with the minimal particle content as given in Fig.~\ref{StandardModel} such that the neutrinos are exactly massless. While neutrino masses can be induced in a minimal way by adding right-handed neutrinos, we consider this as beyond the SM physics.}} required by neutrino oscillations, because in the SM neutrinos are necessarily massless due to the absence of a right-handed partner. 

Unfortunately, no right-handed neutrinos have been observed, Dark Matter direct detection experiments did not see any signal~\cite{LZ:2022lsv}. {Thus, there are many options for how SM can be extended,} spanning a very large mass range (from several keV to the scale of Grand Unification at around $10^{15}$GeV), to account for Dark Matter and neutrino masses. Therefore, more experimental information on physics beyond the SM, preferably deviations from its predictions (in the best case in the form of new resonances) is imperative to make progress towards a theory superseding the SM. 

The SM of particle physics has been extensively tested~\cite{ParticleDataGroup:2022pth} and the search for new physics (NP) {continued} in the last decade, both in high-energy searches (mainly at the LHC) and in precision experiments testing new particles. {High-energy experiments try to produce new particles directly in the form of resonances, i.e.~peaks in kinematic distributions and, thus, are able to measure directly the mass of the new states. Precision experiments look for the effect of new particles via quantum fluctuations and can determine the ratio of the new physics coupling over the mass scales squared but not the mass itself. However, indirect searches for many NP models are able to test higher scales than what can be accessed directly at the LHC (which is limited by a bit lesst than half the center of mass energy, i.e.~$\approx6\,$TeV for the LHC).}

In fact, an increasing number of hints for new physics, i.e.~deviations from the SM predictions called “anomalies”, have been reported. They span over a huge energy range and include precision measurements of muon properties (the anomalous magnetic moment of the muon), over semi-leptonic $B$ meson (quark anti-quark bound states containing a $b$ quark) decays, the measurement of the $W$ boson mass, to direct LHC searches and even non-resonant searches for particles too heavy to be produced directly at the LHC. 

While probability theory and statistics tell us that one cannot expect that all these anomalies will be confirmed in the future, it is also unlikely that all of them are just statistical flukes {or due to experimatal errors}. Therefore, it is important to assess the strengths and weaknesses of these anomalies and to assess to which extensions of the SM they point for predicting other signals for future verification or falsification. In this article, we will review the status of these anomalies, give an overview of how they can be explained by BSM physics and provide an outlook on their future implications. 

\section{Anomalies and New Physics}

{An anomaly is generally defined as a deviation from the common rule. In particle physics, this entails a discrepancy between the experimental data and the corresponding SM prediction. An objective selection criterion is essential to choose which of the many presently existing anomalies should be considered for further scrutiny. Here we employ the following criteria:
\begin{itemize}
\item The combined global statistical significance should be at least three standard deviations ($3\sigma$), after the application of trial factors and look-elsewhere effects. As such, we allow for the combination of several channels or measurements if they are related to the same effective interactions or simplified model parameters.
\item The experimental signature should include more than a single channel (or observable) or be measured (maybe with lower significance) by more than one independent experiment.
\item The deviation should be described by a theoretically robust model that does not contradict the wealth of existing constraints from other measurements in particle physics. 
\end{itemize}

In general, the discovery threshold in particle physics is taken to be $5\sigma$. This means that the probability of a statistical fluctuation is about one in 3.5 million. However, estimating the systematic uncertainty is often very difficult and the probability of a human error is not included in the statistical significance. Therefore, this definition is only applied if the situation is ``clear''.  This means that the measurement is to some extent predicted by a well-established model. This was e.g.~the case for the top quark or the Higgs boson where the SM required their existence. However, for ``unexpected'' measurements one usually requires very high standards concerning the systematic uncertainties (both on the experimental and the theoretical side) and independent confirmation by another experiment or channel.

For direct and indirect searches the analysis of an anomaly proceeds differently. For low-energy precision observables, one can often assume that the scale of new physics is higher than the energy scale of the experiment. One can then evaluate the consistency and significance of several measurements in an effective field theory (EFT) approach. In this setup, the interactions parametrizing the NP effects are required to respect fundamental symmetries like charge conservation and Lorentz invariance. In a later step, one can then assess which NP model can give rise to the desired effective interactions while at the same time avoiding the constraints from other observables which are not directly connected within the effective setup but only arise once the new particle content is known. 

For direct searches, an EFT approach is not possible because one is aiming at the production of new particles. Therefore, they are dynamic degrees of freedom and cannot be integrated out to obtain effective interactions. However, in many cases, one can study what is called ``simplified models''. This means that one does not aim at constructing a UV complete model but only considers adding a single new field with the relevant couplings to the SM. Nonetheless, one can still combine different measurements if the consistency of the suggested masses is given. Furthermore, by assuming specific coupling structures, like that a new Higgs is SM-like, one can reduce the number of free parameters.

Importantly, if any anomaly in direct or indirect searches were confirmed beyond a reasonable doubt, this would inevitably imply the breakdown of the SM and require its extensions by new particles and new interactions.}

\subsection{Extensions of the Standard Model}

{Top-down approaches based on principles like supersymmetry, compositeness or extra-dimensions have not received experimental confirmation so far. Therefore, it might well be that fundamental physics has an unorthodox structure, suggesting that a data-driven bottom-up approach could be more successful.}

For a consistent renormalizable extension of the SM, only scalars bosons (spin 0), fermions (spin 1/2) and vectors bosons (spin 1) are at our disposal, provided that in the latter case a Higgs-like mechanism of spontaneous symmetry-breaking exists to generate their masses. Here we will focus on the following extensions of the SM:

\begin{itemize}
\item Leptoquarks ({LQs}): Scalar or vector bosons that carry color and couple quarks directly to leptons~\cite{Buchmuller:1986zs,Dorsner:2016wpm}. These particles were first proposed in the context of quark-lepton unification at high energies, namely the Pati-Salam model~\cite{Pati:1974yy} and Grand Unified Theories (GUTs)~\cite{Georgi:1974sy}. Furthermore, in the R-parity violating MSSM (see e.g.~Ref.~\cite{Barbier:2004ez} for a review) the supersymmetric partners of quarks can have the properties of LQs.
 \item Di-quarks ({DQs}): Scalar bosons that are either triplets or sextets of $SU(3)_c$ and couple to a quark and an anti-quark. They are predicted by GUTs based on the $E_6$ symmetry group~\cite{Hewett:1988xc} and appear in the R-parity violating MSSM.
 \item $Z^\prime$ bosons: Neutral heavy vector bosons. They can be singlets under $SU(2)_L$ but also the neutral component of an $SU(2)_L$ multiplet. These particles can be resonances of the SM $Z$, e.g.~Kaluza-Klein excitations of the SM $W$ in composite~\cite{Weinberg:1962hj} or extra-dimensional models~\cite{Randall:1999ee}, or originate from an abelian symmetry (like $B-L$~\cite{Pati:1974yy}) or gauged flavour symmetries~\cite{Froggatt:1978nt}.
\item $W^\prime$ bosons: Electrically charged but QCD neutral vector particles. They can have similar origins as $Z^\prime$ bosons but also come for a left-right symmetry~\cite{Mohapatra:1974gc}.
\item Vector-like Quarks ({VLQs}): For vector-like fermions in general left-handed and right-handed fields have the same quantum numbers under the SM gauge group (unlike SM fermions) and can thus have masses independently of electroweak (EW) symmetry breaking, meaning that they can be arbitrarily heavy. They appear in GUTs~\cite{Langacker:1980js}, as resonances of SM fermions in composite or extra-dimensional models~\cite{Antoniadis:1990ew} and as the supersymmetric partners of SM vectors and scalars~\cite{Haber:1984rc}. 
\item Vector-like Leptons ({VLLs}):
 These particles can have similar origins as VLQs. In addition, they are involved in the type~I~\cite{Minkowski:1977sc,Lee:1977tib} and type~III~\cite{Foot:1988aq} seesaw mechanisms used for giving masses to the light active neutrinos as required by neutrino oscillations. 
\item New scalars ($S$): Scalars could be supersymmetric partners of SM fermions~\cite{Haber:1984rc}, but also scalar fields of different representations under the SM gauge group can be added. Most commonly, a copy of the SM Higgs, an $SU(2)_L$ doublet with hypercharge $0$, leading to a two-Higgs doublet model~\cite{Chanowitz:1985ug,Branco:2011iw}. Note that we do not include coloured scalars with the properties of DQs or LQs here.
\item Heavy gluons ($G^\prime$): Heavy gluons are similar to $Z^\prime$ bosons but charged under QCD. They can arise from the breaking of a larger gauge group down to $SU(3)_c$ or be excitations of the SM gluons. 
\end{itemize}

\section{Status, Explanations and Prospects of the Anomalies}

Let us now review the status of these anomalies, how they can be explained by NP and what the future prospects are. We will present them in increasing order of the corresponding energy scale.

\subsection{Anomalous magnetic moment of the muon ($a_\mu$)}

While the Dirac equation predicts that the {gyromagnetic-ratio (the $g$-factor)} of any fundamental fermion is exactly 2, the famous prediction of Quantum Electrodynamics (the quantum field theory of electromagnetism) by Schwinger~\cite{Schwinger:1948iu} was a positive shift of $a_\ell=(g-2)_\ell/2=\alpha/(2\pi)$ (see left diagram in Fig.~\ref{FeynmanDiagrams} a)). Nowadays, the accuracy has dramatically increased. The combined value of the 2006 result of the Brookhaven E821 experiment~\cite{Bennett:2006fi} and the {recent} $g-2$ experiment at Fermilab~\cite{Abi:2021gix,Muong-2:2023cdq} deviate from the SM prediction of the $g-2$ theory initiative~\cite{Aoyama:2019ryr} by $5.1\,\sigma$. However, this SM prediction is based on measurements of $e^+e^-\to$hadrons\footnote{Hadrons are bound states of quarks which form due to the confining nature of QCD (i.e.~because the coupling increases at low energies).}~\cite{Colangelo:2018mtw,Davier:2019can,Keshavarzi:2019abf} and does not include newer results from lattice simulations of 
Quantum Chromodynamics (the quantum field theory describing the strong interactions)~\cite{Borsanyi:2020mff} nor the latest measurement of $e^+e^-\to$hadrons by the CMD~3 collaboration~\cite{CMD-3:2023alj} which would render the SM prediction closer to the measurement~\cite{Colangelo:2023rqr}. Since these tensions between the different SM predictions are not understood yet, one can only say that a positive shift in $a_\mu$ of the order of $10^{-9}$ is preferred but a reliable estimate of the significance is not possible at the moment.

Since the anomalous magnetic moment of the electron is measured and predicted much more precisely than $a_\mu$, the resulting bounds on NP are stringent~\cite{Hanneke:2008tm,Aoyama:2017uqe,Laporta:2017okg}. Therefore, the effect in $a_e$ must be smaller than in $a_\mu$, unlike the famous Schwinger term, and thus violate lepton flavour universality~\cite{Crivellin:2018qmi} (see Ref.~\cite{Athron:2021iuf} for a recent overview on NP in $a_\mu$). Furthermore, because the deviation from the SM prediction is as large as its EW contribution, new physics must be quite light, e.g.~a light $Z^\prime$ boson coupling {to muons but not to electrons with a mass below the muon threshold ($\approx$0.2\,GeV) to avoid limits from $e^+e^-\to 4\mu$ from Babar~\cite{BaBar:2016sci} and Belle~\cite{Belle:2021feg} is a viable option~\cite{Ma:2001md,Baek:2001kca}.} Alternatively, if NP is heavy (at the TeV scale) it must possess an enhancement factor. This can be provided via the mechanism of chiral enhancement, meaning that the chirality flip does not originate from the small muon Yukawa coupling (like for the SM contribution to $a_\mu$) but from a larger coupling of other particles to the SM Higgs. In the MSSM, this factor is $\tan\beta$, the ratio of the two vacuum expectation values of the two Higgs fields~\cite{Everett:2001tq,Feng:2001tr} but also models with generic new scalars and fermions can explain $a_\mu$~\cite{Czarnecki:2001pv,Kannike:2011ng,Kowalska:2017iqv,Crivellin:2021rbq}. Furthermore, there are two scalar LQs ($S_1$ and $S_2$) that address $a_\mu$ via a $m_t/m_\mu$ enhancement~\cite{Djouadi:1989md,Bauer:2015knc,Crivellin:2018qmi} (see Fig.~\ref{FeynmanDiagrams} a)). This top mass enhancement also leads to interesting predictions for $h\to\mu^+\mu^-$~\cite{Crivellin:2020tsz} and $Z\to\mu^+\mu^-$~\cite{ColuccioLeskow:2016dox} that can be measured at future colliders.

While the experimental situation concerning the direct measurement seems settled, there will be updates on $e^+e^-\to$hadrons e.g.~by Belle-II~\cite{Belle-II:2018jsg} and the MUonE~\cite{Abbiendi:2016xup} experiment aims at an independent determination of the disputed SM contribution with a completely different method. Furthermore, lattice QCD simulations will deliver improved results within the next years. While the anomalous magnetic moment of the tau is in principle very sensitive to NP,  its measurement is very difficult. In fact, recent determinations are not even sensitive to the Schwinger term~\cite{ATLAS:2022ryk,Haisch:2023upo} and the currently only realistic option for reaching a sensitivity that is constraining NP seems to be $\tau$ pair production at Belle-II with polarized beams~\cite{Crivellin:2021spu}.

\subsection{The 17\,MeV Anomaly in excited nuclei decays (${X17}$)}

An experiment conducted at the Atomki laboratory (Debrecen, Hungary) studied the nuclear reaction
$^{7}$Li(p,e$^{+}$e$^{-})^{8}$Be~\cite{Krasznahorkay:2015iga}. The experiment consisted of a proton capture process of Lithium and measuring the relative angle between the positron-electron pair emitted in the decay of the excited state to the ground state of Beryllium. Subsequently, similar excesses also emerged in the decays of excited $^4$He and $^{12}$C nuclei~\cite{Krasznahorkay:2021joi,Krasznahorkay:2022pxs}. The statistical significance exceeds $6\sigma$ in all cases. While previously similar excesses~\cite{deBoer:1996qdk} measured by the same collaboration disappeared later~\cite{deBoer:2001sjo}, the current one has been checked with different experimental setups, position-sensitive detectors, and varying beam energies and appears at different angles with different target nuclei. However, the possibility of a SM effect is still not excluded~\cite{Aleksejevs:2021zjw}. A recent review of experimental and theoretical aspects of the anomaly can be found in Ref.~\cite{Alves:2023ree}.

Concerning NP explanation, a particle with a mass of $\approx17\,$MeV is consistent with the angular measurements of both $^{8}$Be and $^4$He. From parity consideration (in the case of CP conservation) only vector, axial-vector or pseudo-scalar states are possible. Interestingly, it is possible that the hypothetical $X17$ boson could be related to $g-2$ of the muon~\cite{Nomura:2020kcw} or the neutron lifetime puzzle~\cite{TienDu:2020wks} (to be discussed next). 

There is an effort to explore the $X17$ anomaly at various facilities. The resonant production of  $X17$ with an electron-positron beam at the Positron Annihilation into Dark Matter Experiment (PADME) in Frascati has been discussed~\cite{Darme:2022zfw}. A search for a new vector boson in the same mass range is planned at the Mu3e experiment at the Paul Scherrer Institute~\cite{Echenard:2014lma}. 
The New Judicious Experiments for Dark Sector Investigations (New JEDI) project has launched experiments at the ANDROMEDE facility at Orsay, and the iThemba LABS accelerator and the half-AFRODITE detectors in Cape Town also plan investigations of $X17$~\cite{Bastin:2023utm}. Measurements are also possible at other facilities, such as CERN, JLAB, Novosibirsk and Mainz~\cite{Alves:2023ree}.

\subsection{Anomalies in electron neutrino appearance and disappearance ($\nu_e$)}

Neutrino appearance anomalies revolve around excesses in the quasi-elastic production of electrons from accelerator neutrinos reported by LSND~\cite{LSND:2001aii} and MiniBooNE~\cite{MiniBooNE:2018esg}. In particular, the MiniBooNE experiment observed an excess in the neutrino energy range $200<E_\nu<1250$\,MeV with a significance of 4.8$\sigma$~\cite{MiniBooNE:2020pnu}. However, a reanalysis of the theoretical uncertainties leads to a smaller tension with the SM prediction~\cite{Brdar:2021ysi}. The MicroBooNE experiment began operating in 2015~\cite{MicroBooNE:2016pwy} and results indicate that the MiniBooNE excess cannot be explained entirely by electron neutrino appearance~\cite{MicroBooNE:2021tya}. 

A first combined analysis of the MiniBooNE and MicroBooNE data within the SM plus a single sterile neutrino\footnote{A sterile neutrino is a fermion which is a singlet under the SM gauge group and mixes with the SM neutrino via a Yukawa coupling to the SM Higgs.} finds a preference over the SM of 4.6$\sigma$ for MiniBooNE alone while the addition of the MicroBooNE exclusive (inclusive) electron-neutrino data reduces the significance to 4.3$\sigma$ (3.4$\sigma$)~\cite{MiniBooNE:2022emn}. Furthermore, once constraints from MINOS~\cite{MINOS:2017cae} and IceCube~\cite{IceCube:2020phf} are included, the preference for sterile-active neutrino mixing is further diminished~\cite{Dentler:2018sju}. Therefore, more exotic NP options like energy-dependent mixing parameters have been considered~\cite{Babu:2022non} (see Ref.~\cite{Acero:2022wqg} for a review).

Anomalies suggesting the disappearance of $\nu_e$ and $\overline{\nu}_e$ are observed in reactor neutrino experiments~\cite{Declais:1994su,CHOOZ:2002qts,Mention:2011rk} 
and the Gallium radioactive source experiments GALLEX~\cite{GALLEX:1997lja,Kaether:2010ag} and SAGE~\cite{SAGE:2009eeu}. The latter experiments use intense $^{51}$Cr and $^{37}$Ar radioactive sources inside the detectors and search for the inverse beta reaction $\nu_e+^{71}$Ga$\rightarrow ^{71}$Ge$+e^-$. The results indicate less $^{71}$Ge occurrences than expected. The significance of the deficit is about 5$\sigma$ and is referred to as the Gallium anomaly~\cite{Acero:2007su,Giunti:2010zu}. Ref.~\cite{Brdar:2023cms} critically evaluated the assumptions underlying the Gallium anomaly and found several possible caveats (like the Gallium lifetime or neutrino flux) that could account for the excess without the involvement of NP. 

A straightforward explanation in terms of active-sterile neutrino oscillations is excluded by solar and reactor neutrino experiments~\cite{Giunti:2021kab,Berryman:2021yan}. Therefore, one again has to resort to more exotic options like a parametric resonance~\cite{Losada:2022uvr} (see Ref.~\cite{Brdar:2023cms}) for a review.

The proposed explanations of the Gallium anomalies
can be tested by measurements with a different neutrino source, like $^{65}$Zn at the BEST experiment~\cite{Barinov:2022wfh}, 
and/or other detection material (e.g.~$^{37}$Cl). Furthermore, DUNE~\cite{DUNE:2015lol} and the Short Baseline Neutrino program at Fermilab~\cite{Fava:2019fuz} will scrutinize our understanding of neutrino oscillations.

\subsection{$\beta$-Decay Anomalies ($\beta$)}

{The CKM matrix $V$ is by construction unitary~\cite{Kobayashi:1973fv}, therefore, as required, conserving probability.} We know from experiments that it has a hierarchical structure; while the size of the diagonal elements is close to unity, the off-diagonal elements are small. One can test the SM prediction $\sum_j V_{ij}V^*_{jk}=\delta_{ik}$ experimentally. In this context, the Cabibbo angle~\cite{Cabibbo:1963yz}, which parametrizes the mixing between the first two quark generations, is particularly interesting as it dominates the first and second row and column relations. In fact, a deficit in first-row and first-column CKM unitarity, which can be traced back to the fact that $V_{ud}$ extracted from beta decays~\cite{Czarnecki:2018okw,Hardy:2020qwl} (see left diagram in Fig.~\ref{FeynmanDiagrams} b)) does not agree with $V_{us}$ ($V_{cd}$) determined from kaon\footnote{Kaons are made of a light anti-quark (up or down) and a strange quark. A $\phi$ is a strange anti-strange bound state and a ``$*$'' refers to an excited bound state.} and tau decays ($D$ decays), when comparing them via CKM unitarity. Furthermore, there is also a disagreement between the determinations of $V_{us}$ from $K\to\mu\nu$~\cite{KLOE:2005xes} and $K\to \pi\ell\nu$~\cite{KLOE:2007wlh} decays. The significance of these deviations crucially depends on the radiative (quantum) corrections to beta decays~\cite{Marciano:2005ec,Seng:2018yzq,Hardy:2020qwl} and on the treatment of the tensions between kaon~\cite{Moulson:2017ive,Seng:2021nar,Cirigliano:2022yyo} and tau decays~\cite{HFLAV:2022pwe}. In summary, both tensions are slightly below the $3\sigma$ level.

A sub-permille effect suffices to explain these tensions. The disagreement between the two determinations of $V_{us}$ can only be explained via a right-handed quark current pointing towards VLQs (see Fig.~\ref{FeynmanDiagrams} b)). The deficit in first-row and first-column CKM unitarity can be explained via left-handed (i.e.~SM-like) NP in beta decays. Both an effect in beta decays or in the Fermi constant (determined from muon decay, i.e.~$\mu\to e\nu\nu$), which is needed to extract $V_{ud}$, is possible. Therefore, we have four options~\cite{Crivellin:2021njn}: 1) a direct (tree-level) modification of beta decays 2) a direct (tree-level) modification of muon decay 3) a modified $W$-$\mu$-$\nu$ coupling entering muon decay 4) a modified $W$-$u$-$d$ coupling entering beta decays (the effect of a modified $W$-$w$-$\nu$ drops out). Option 1) could in principle be realized by a $W^\prime$~\cite{Capdevila:2020rrl} or a LQ~\cite{Crivellin:2021egp}, however, in the latter case, stringent bounds from other flavour observables arise. Possibility 2) can be achieved by adding a singly charged $SU(2)_L$ singlet scalar~\cite{Crivellin:2020klg}, a $W^\prime$~\cite{Capdevila:2020rrl} or $Z^\prime$ boson with flavour violating couplings~\cite{Buras:2021btx}. Option 3) and 4) can be achieved by vector-like leptons~\cite{Coutinho:2019aiy,Kirk:2020wdk} and VLQs~\cite{Belfatto:2019swo,Branco:2021vhs,Belfatto:2021jhf,Crivellin:2022rhw}, respectively. However, without a compensating effect~\cite{Crivellin:2020ebi}, explaining the CAA via a modification of $G_F$ increases the tension in the $W$ mass. 

There is also a significant tension ($\approx 4\sigma$) between the neutron lifetime (and thus $V_{ud}$) determined from beam and bottle experiments. The average of the beam values is $\tau_n=888.2 \pm 2.0 \mathrm{~s}$~\cite{Byrne:1996zz,Yue:2013qrc} while the best determination from bottle experiments is $\tau_n=877.75 \pm 0.28_{\text {stat }}+0.22 /-0.16_{\text {syst }} \mathrm{s}$~\cite{UCNt:2021pcg}. 

To understand possible NP explanations, it is important to note that while in bottle experiments the remaining neutrons are counted, beam experiments count the decay protons. Therefore, if the branching ratio of neutrons to final states with protons is not 100\%, the lifetime measured in beam experiments would be larger than the real neutron lifetime. However, the beam/bottle discrepancy cannot be explained within an effective field theory setup, i.e.~with heavy NP since this would lead to proton decay which is very tightly constrained. However, neutron decay to light dark matter particles could result in a stable proton for very fine-tuned mass configurations~\cite{Fornal:2018eol}. Alternatively, neutron oscillations into mirror neutrons have been proposed as an explanation~\cite{Berezhiani:2018eds}.

Improved measurements and theory calculations of beta decays will be available within the next years~\cite{Brodeur:2023eul}. Furthermore, NA62 could measure $(K\to \mu\nu)/(K\to \pi \mu\nu)$ to asses the possibility of right-handed currents~\cite{Cirigliano:2022yyo} and the PIONEER experiment will measure pion beta decay~\cite{PIONEER:2022yag} to determine $V_{us}$, which is theoretically accurately predicted. The neutron anti-neutron explanation of the lifetime puzzle can be tested at the PSI ultracold neutron source~\cite{Ayres:2021zbh}.

\subsection{Hadronic meson decays ($M\to mm^\prime$)}
Several anomalies in the decay of mesons into two lighter ones have been observed. This includes $CP$-violation\footnote{$CP$ is the combined application of charge conjugation in parity transformation (mirroring). In the SM $CP$ is conserved by all interactions except transitions involving the single complex phase phase of the CKM matrix. Therefore, $CP$ is only induced via the $W$ boson and thus very small within the SM.} in $b\to s$ transitions and $D$ meson decays, which is known to be very suppressed in the SM, and total branching for charged current $B$ decays.

The improved SM predictions for the total branching ratios $\bar B\to D^{(*)}K$ and $\bar B_s\to D^{(*)}_s\pi$~\cite{Bordone:2020gao} based on QCD factorization~\cite{Beneke:2001ev} deviate from the corresponding measurements~\cite{ParticleDataGroup:2022pth} by a combined significance of $5.6\sigma$. However, since the anomaly is observed in total rates no cancellation or suppression of QCD uncertainties occurs and Ref.~\cite{Piscopo:2023opf} challenged the accuracy of the SM values. Since these are charged current processes, i.e.~mediated at tree-level in the SM, a NP explanation is particularly challenging taking into account that a $O(50\%)$ effect is needed. Therefore, both $W^\prime$ models~\cite{Iguro:2020ndk} and di-quark explanations are stringently constrained by LHC searches~\cite{Bordone:2021cca}.

The first evidence for CP violation in $D$ decays was observed in 2011 by LHCb~\cite{LHCb:2011osy} and the discovery level was reached in 2019 with the differences of CP asymmetries between $D\to KK$ and $D\to \pi^+\pi^-$ given by $\Delta A_{CP}^{\rm LHCb}=(-15.4 \pm 2.9) \times 10^{-4}$~\cite{LHCb:2019hro}. This has to be compared to the estimate of the SM prediction, which is notoriously difficult for charm physics, of $|\Delta A_{CP}^{\rm LHCb}|<3.6 \times 10^{-4}$~\cite{Chala:2019fdb}. Furthermore,  Ref.~\cite{LHCb:2022lry} determined directly the CP asymmetry in $D\to K^+K^-$ 
which allows for a test of $U$-spin symmetry\footnote{Isospin symmetry used the fact that for the strong interactions up and down quarks are too a good approximation indistinguishable as they are very light. For $U$-spin symmetry, this concept is extended to include the strange quarks. However, since the strange is much more massive than first-generation quarks, the symmetry is more strongly broken and the resulting predictions are less reliable. } of the SM and shows indications of a violation of it~\cite{Bause:2022jes}. 

An overview of NP explanations of CP asymmetries in $D$ decays was already given in Ref.~\cite{Altmannshofer:2012ur}, including $Z^\prime$ bosons and di-quarks. Already at that time, the solutions were under pressure from LHC searches and the corresponding limits got much more stringent in the meantime.

There are also hints of BSM CP violation in hadronic $B$ meson decays with $b\to s$ transitions. This includes the long-standing $B\to K\pi$ puzzle~\cite{Buras:2003dj} which was confirmed by LHCb~\cite{LHCb:2020dpr}. While here the theory predictions within the SM seem to be more reliable as they use isospin relations~\cite{Fleischer:2017vrb} the significance is around $3\sigma$ but supported by $B_s\to KK$ measurements~\cite{LHCb:2018pff}. Finally, even though not CP-violating, there are indications of $U$-spin violation in polarization observables~\cite{Alguero:2020xca}. For explaining these hints for BSM $CP$ violation, somehow smaller NP effects are required than for the total branching ratios or the polarization observables, making an explanation via $Z^\prime$ bosons or heavy gluons easier but still not straightforward~\cite{Calibbi:2019lvs} while the constraints on di-quark models are expected to be less stringent. 

Progress on the SM side does not seem easy. However, a lot of forthcoming data is expected by Belle-II and LHCb. Furthermore, the hints for $CP$ violation in $B$ and $D$ decays could be related to direct $CP$ violation in the Kaon system ($\epsilon^\prime/\epsilon$)~\cite{Crivellin:2019isj} (see Ref.~\cite{Buras:2022cyc} for an overview on the recent status of $\epsilon^\prime/\epsilon$) and $U$-spin violation~\cite{Bhattacharya:2022akr}. 

\subsection{Charged current tauonic $B$ decays ($R(D^{(*)})$)}

These charged current transitions, mediated at tree-level by a $W$ boson in the SM (see left diagram in Fig.~\ref{FeynmanDiagrams} c)), have significant branching ratios (up to ${\mathcal O}(10^{-2})$). With light leptons, they are used to extract the CKM element $V_{cb}$\footnote{Note that there is also a long-lasting tension in the inclusive vs exclusive determination of $V_{cb}$~\cite{Gambino:2019sif} (and $V_{ub}$) where $B \to {D^{\left( * \right)}}\ell \nu$ is involved. However, it has been shown that this anomaly cannot be explained by NP~\cite{Crivellin:2014zpa}. } and the result is consistent with the global CKM fit~\cite{Charles:2004jd,UTfit:2005ras}. However, the ratios (of branching ratios) $R( {{D^{\left( * \right)}}})$ = ${\rm Br}({B \to {D^{\left( * \right)}}\tau \nu })$/ ${\rm Br}({ B \to {D^{\left( * \right)}}\ell \nu )})$, are measured to be bigger than the SM predictions by approximately 20\%, resulting in a $\gtrapprox3\sigma$ significance~\cite{HFLAV:2022pwe} for NP related to tau leptons.\footnote{The analysis by BaBar~\cite{BaBar:2012obs,BaBar:2013mob}, Belle~\cite{Belle:2015qfa,Belle:2016ure,Hirose:2016wfn,Hirose:2017dxl, Belle:2019rba} and LHCb~\cite{Aaij:2015yra,Aaij:2017uff,Aaij:2017deq} used different tag and tau reconstruction methods. The interested reader is referred to the online update of Ref.~\cite{HFLAV:2022pwe} for an overview.} 

This transition occurs at tree-level in the SM. Therefore, also a tree-level NP effect is necessary to obtain the needed effect of $O(10)$\% w.r.t.~the SM (assuming heavy NP with perturbative couplings). Therefore, charged Higgses~\cite{Crivellin:2012ye,Fajfer:2012jt,Celis:2012dk}, $W’$~bosons~\cite{Bhattacharya:2014wla} (with or without right-handed neutrinos) or LQs~\cite{Sakaki:2013bfa,Bauer:2015knc,Freytsis:2015qca,Fajfer:2015ycq} are candidates. While there is a small region in parameter space left that can account for $R(D^{(*)})$ with charged Higgses~\cite{Iguro:2022uzz,Blanke:2022pjy}, LHC searches constrain $W^\prime$ solutions~\cite{Bhattacharya:2014wla,Greljo:2015mma}, leaving LQ as the probably best solutions (see Fig.~\ref{FeynmanDiagrams} c)). However, also for LQs constraints from $B_s-\bar B_s$ mixing, $B\to K^{(*)}\nu\nu$ and LHC searches must be respected such that the $SU(2)_L$ singlet vector LQ~\cite{Calibbi:2015kma,Barbieri:2016las,DiLuzio:2017vat,Calibbi:2017qbu,Bordone:2017bld,Blanke:2018sro,King:2021jeo} or the singlet-triplet model~\cite{Crivellin:2017zlb,Crivellin:2019dwb,Gherardi:2020qhc} are particularly interesting. 

Concerning future prospects, $R(D^{(*)})$ and related ratios can be measured at Belle-II~\cite{Belle-II:2018jsg}, by LHCb with Run~3 data and the parked $B$ data from CMS~\cite{Bainbridge:2020pgi}. Furthermore, polarization observables will be measured so precisely that they can distinguish between different NP models and an improvement in the form-factors from lattice QCD is expected.

\subsection{Flavour changing neutral current semi-leptonic $B$ decays ($b\to s\ell^+\ell^-$)}

Like all flavour changing neutral current processes, $b\to s\ell^+\ell^-$ transitions are loop suppressed {(i.e.~by the small probability that two particles are produced via quantum fluctuations and annihilate again)} within the SM (see left diagram in Fig.~\ref{FeynmanDiagrams} d) for an example) since only the couplings of the charged $W$ can violate quark flavour. This results in small branching ratios, up to a few times $10^{-6}$. While the previous hints~\cite{LHCb:2021lvy} for lepton flavour universality violation in the ratios 
$R({K^{\left(*\right)}})  = {\rm Br}({B \to K^{\left( * \right)}\mu ^+\mu ^-})
/{\rm Br}({B \to K^{\left( * \right)}e^+e^-})$
were not confirmed~\cite{LHCb:2022qnv} and $B_s\to\mu^+\mu^-$~\cite{LHCb:2020zud,CMS:2022mgd} now agrees quite well with the SM prediction~\cite{Hermann:2013kca,Beneke:2017vpq}, there are several $b \to s \mu^+\mu^-$ observables that significantly deviate from the SM predictions. This includes the angular observable $P_5^{\prime}$~\cite{Descotes-Genon:2012isb,LHCb:2020lmf}, the total branching ratio ${\rm Br}({B \to K\mu^+\mu^-})$\cite{LHCb:2014cxe,Parrott:2022zte}, ${\rm Br}({B_s \to \phi\mu^+\mu^-})$~\cite{LHCb:2021zwz,Gubernari:2022hxn} and also semi-inclusive observables~\cite{Isidori:2023unk}. As a result, global fits find a preference for NP at the $5\sigma$ level~\cite{Buras:2022qip,Ciuchini:2022wbq,Alguero:2023jeh}. Recently, the Belle-II collaboration reported an {2.8$\sigma$ excess over the SM hypothesis} in the closely related $B\to K^*\nu\bar \nu$ decay~\cite{BelleIISem}.

The new measurements of $R({K^{\left(*\right)}})$ require dominantly lepton flavour universal NP and $B_s\to \mu\mu$ constrains axial couplings to leptons. Such a NP effect at the required level of $O(20\%)$ (w.r.t.~the SM) can be most naturally obtained via~\cite{Alguero:2022wkd}: 1)~A $Z’$ boson with lepton flavour universal but flavour-violating couplings to bottom and strange quarks~\cite{Buras:2013qja,Gauld:2013qba} (see Fig.~\ref{FeynmanDiagrams} d)). However, due to the bounds from $B_s-\bar B_s$ mixing~\cite{DiLuzio:2017fdq}, the LHC (see e.g.~\cite{Allanach:2015gkd}), and LEP~\cite{Electroweak:2003ram} a full explanation requires some tuning in $B_s-\bar B_s$ mixing by a right-handed $sb$ coupling~\cite{Crivellin:2015era} or a cancellation with Higgs contributions~\cite{Crivellin:2015lwa}. Furthermore, $K^0-\bar K^0$ and $D^0-\bar D^0$ mixing require an approximate global $U(2)$ flavour symmetry~\cite{Calibbi:2019lvs}. 2)~$\tau$ or charm loop effect via an off-shell photon penguin~\cite{Bobeth:2014rda}. The LQ representations which can give such a tau loop are $S_2$ LQ~\cite{Crivellin:2022mff}, the $U_1$ LQ~\cite{Crivellin:2018yvo} or the combination of $S_1+S_3$~\cite{Crivellin:2019dwb}. The 2HDM with generic flavour structure~\cite{Crivellin:2013wna} can generate the desired effect via a charm loop $C_9^U$~\cite{Crivellin:2019dun,Iguro:2023jju}. Alternatively, a DQ solution is possible~\cite{Crivellin:2023saq}.

The best hope to solve the bottleneck concerning the SM predictions is to improve lattice calculations over the full $q^2$ range, like performed in Ref.~\cite{Parrott:2022zte}, and to combine them with other non-perturbative methods like dispersion relations~\cite{Gubernari:2023puw}. Or the experimental side, again Belle-II, LHCb and the parked $B$ program of CMS will help to resolve the situation.

\subsection{$W$ boson mass ($m_W$)}

In general, three parameters are sufficient to parameterise completely (at tree level) the EW sector of the SM. They are usually taken to be the Fermi constant $G_F$, the fine-structure constant $\alpha$ and the $Z$ boson mass since these are measured most precisely. In this input scheme, the $W$ mass is not a free parameter but can be calculated as a function of $G_F$, $\alpha$ and $m_Z$ (and the Higgs and the top mass which enter at the loop-level). The CDF~II result~\cite{CDF:2022hxs} shows a very strong $7\sigma$ tension with the SM prediction. However, LHC~\cite{ATLAS:2017rzl,CMS:2011utm,LHCb:2015jyu,LHCb:2021bjt} and LEP results~\cite{ALEPH:2013dgf} are closer to the SM {with a tension of only 1.8$\sigma$} and thus in tension with CDF~II value. Therefore, employing a conservative error estimate, {following the Particle Data Group recommendation~\cite{ParticleDataGroup:2022pth}}, that increases the uncertainty finds a tension of $3.7\,\sigma$~\cite{deBlas:2022hdk}.\footnote{This average does not include the latest ATLAS result~\cite{ATLAS:2023fsi} superseding Ref.~\cite{ATLAS:2017rzl}, which however has a small impact on the fit.} 
Within the EW fit, there are also tensions due to the forward-backward asymmetry measurement in $Z\to b\bar b$~\cite{ALEPH:2005ab} ($\approx2\sigma$) and in the lepton asymmetry parameter $A_\ell$~\cite{deBlas:2022hdk}, mainly due to the electron channel. 

The tension in the $W$ mass is most easily explained by a tree-level effect, e.g.~an $SU(2)_L$ scalar triplet that acquires a vacuum expectation value~\cite{Konetschny:1977bn} or via $Z-Z^\prime$ mixing (see Fig.~\ref{FeynmanDiagrams} e) in case $Z^\prime$ is an $SU(2)_L$ singlet~\cite{Alguero:2022est}. However, loop effects of new particles with masses below or at the TeV scale are possible as well~\cite{Strumia:2022qkt}, such as VLQs~\cite{Crivellin:2022fdf} or LQs~\cite{Crivellin:2020ukd}.

Since with increased instantaneous luminosity, like at the high-luminosity LHC, the measurement of the $W$ mass at a hadron collider becomes more difficult. However, very precise results would be possible with a future electron-positron collider like ILC~\cite{ILC:2013jhg}, CLIC~\cite{Linssen:2012hp,CLICdp:2018cto}, FCC-ee~\cite{FCC:2018evy,FCC:2018byv} or CEPC~\cite{CEPCStudyGroup:2018ghi,An:2018dwb} which would also improve significantly on the precision of the input parameters for the EW fit. Nonetheless, LHCb could help to solve the puzzle since it does not use the full LHC luminosity.

\begin{figure*}
\begin{center}
	\vspace{-3mm}
	\;\;\qquad\qquad SM \qquad\qquad\qquad\qquad\qquad NP \qquad\qquad\qquad\qquad\qquad SM \qquad\qquad\qquad\qquad\qquad NP\newline \includegraphics[width=0.54\textwidth]{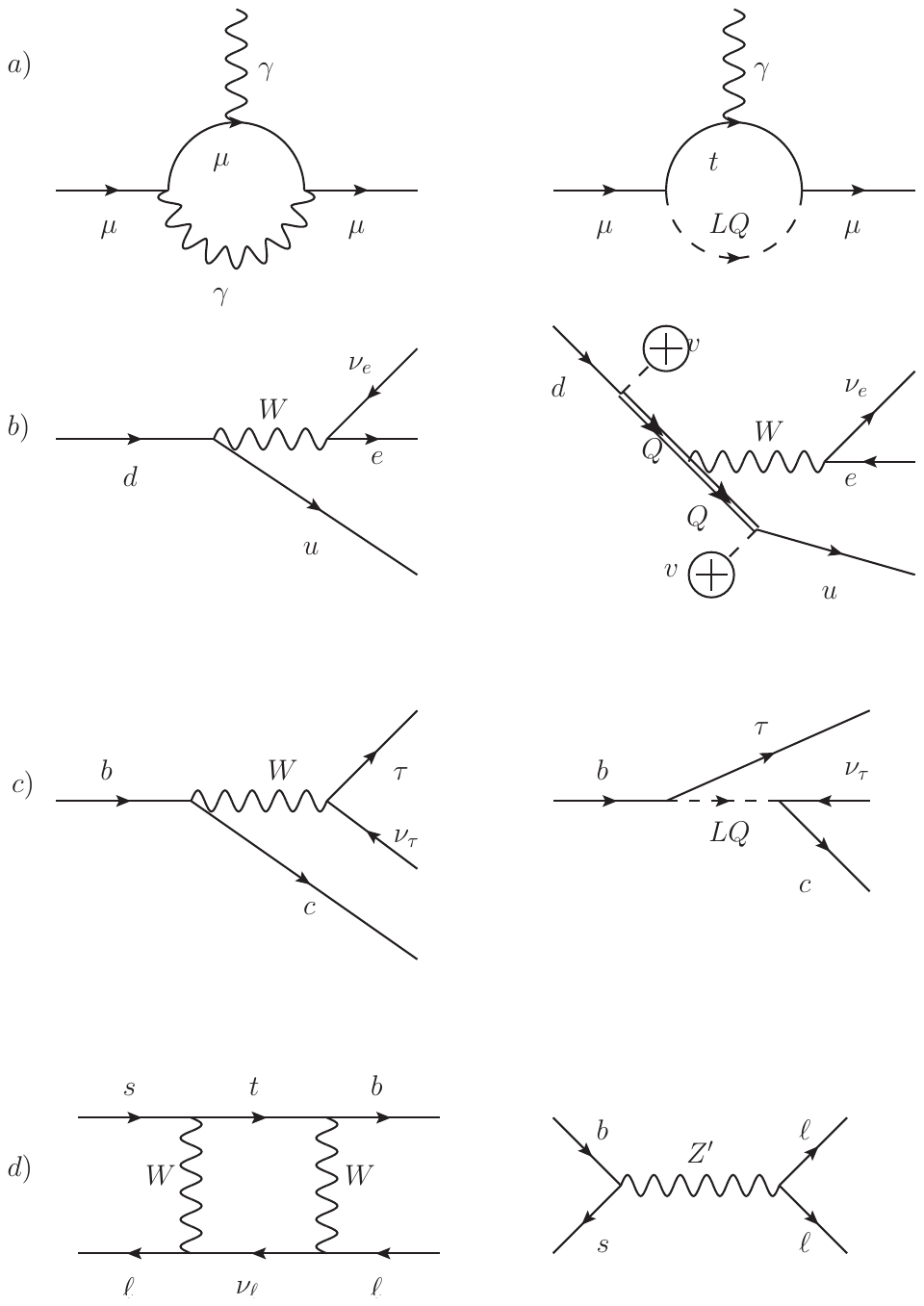}\hspace{-15mm}
 	\includegraphics[width=0.54\textwidth]{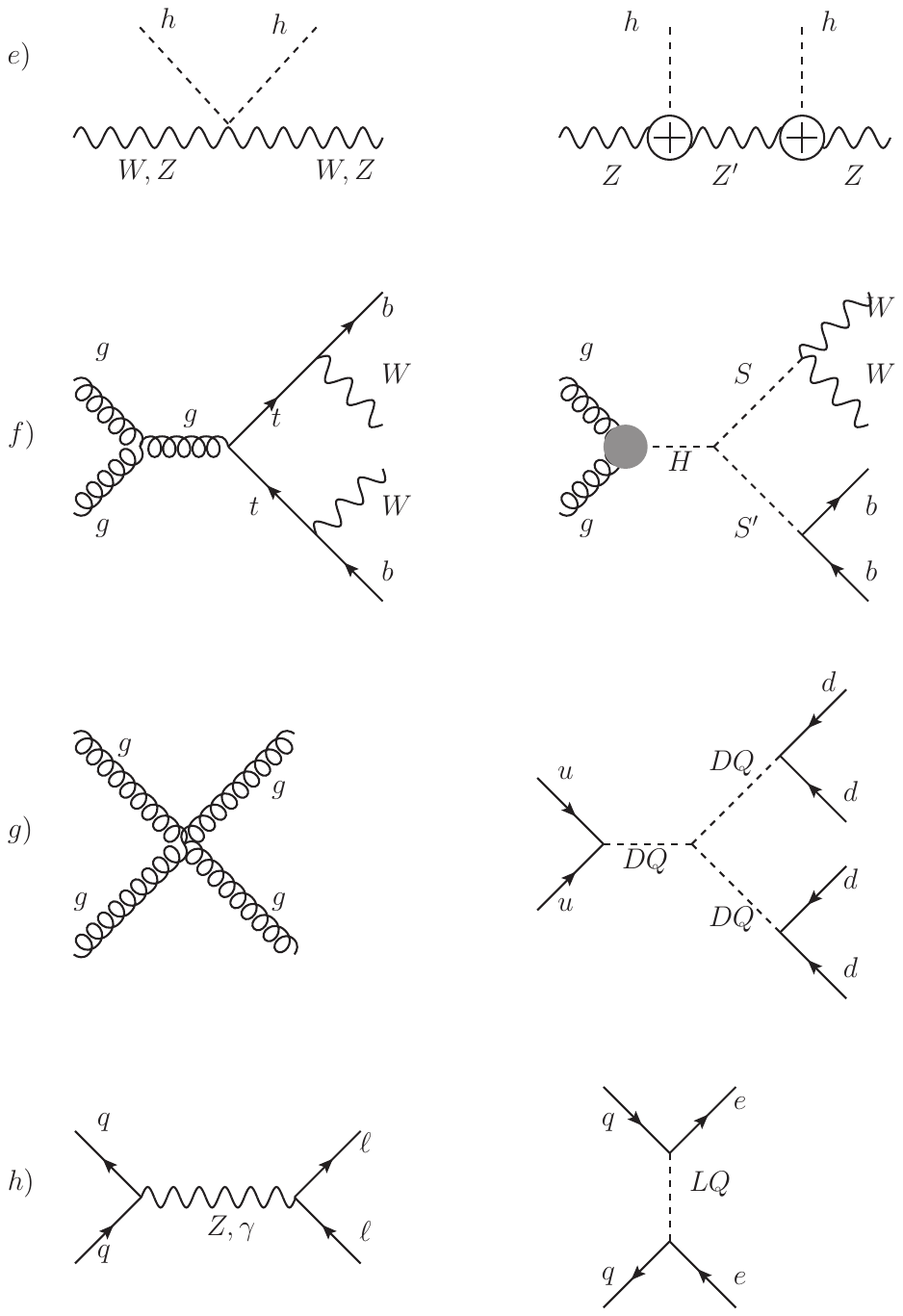}
	\vspace{-10mm} 
\end{center}
\caption{Feynman diagrams showing some of the processes where anomalies are observed. The left diagrams depict the SM process, while the right-handed ones show a possible NP explanation. $\left.a\right)$ Schwinger term contribution to $a_\mu$ and LQ explanation $\left.b\right)$ Leading $\beta$-decay contribution in the SM and modification via a vector-like quark $\left.c\right)$ $W$ contribution to $R(D^{(*)})$ and LQ effect $\left.d\right)$ $W$ box contribution to $b\to s\ell^+\ell^-$ in the SM and $Z^\prime$ effect $\left.e\right)$ $Z\to b\bar b$ and its modification via vector-like quarks $\left.f\right)$ top pair production and decay in the SM and new Higgses ``polluting'' the measurement $\left.g\right)$ di-di-jet production in the SM and NP contribution via DQs h$)$ $pp\to e^+e^-$ in the SM and LQ contribution. }
\label{FeynmanDiagrams}
\end{figure*}

\subsection{LHC Multi-Lepton Anomalies ($e\mu(+b)$)}

The ``multi-lepton anomalies'', are LHC processes with two or more leptons in the final state (see Ref.~\cite{Fischer:2021sqw} for a review), with and without $b$-jets\footnote{Since quarks and gluons are confined at low energies, they do not appear as free particles in a detector but rather hadronize and give signatures called jets ($j$). A $b$-jet is such an experimental signature containing a bottom quark.}, where statistically significant disagreements with the SM predictions have been observed~\cite{vonBuddenbrock:2016rmr,vonBuddenbrock:2017gvy,Buddenbrock:2019tua,Hernandez:2019geu,vonBuddenbrock:2020ter,Banik:2023vxa}. Some of the excesses already emerged with run-1 data (2011-2012) of ATLAS and CMS~\cite{vonBuddenbrock:2016rmr,vonBuddenbrock:2017gvy}. They were confirmed by independent and larger data sets taken during run~2~\cite{Buddenbrock:2019tua,Hernandez:2019geu,ATLAS:2023gsl} leading to disagreements with the SM, exceeding the $5\sigma$ threshold:
\begin{center}
      \begin{tabular}{c|c|c|c}
Final state & Characteristics & SM backgrounds & Significance \\
\hline
$\ell^+\ell^-$+($b$-jets)~\cite{Buddenbrock:2019tua,ATLAS:2023gsl,Banik:2023vxa} & $m_{\ell\ell}<100$\,GeV & $t\overline{t}, Wt$ & $>5\sigma$ \\ 
$\ell^+\ell^-$+(no jet)~\cite{vonBuddenbrock:2017gvy,ATLAS:2019rob} & $m_{\ell\ell}<100$\,GeV & $W^+W^-$ & $\approx 3\sigma$ \\ 
$\ell^\pm\ell^\pm, 3\ell$ + ($b$-jets)~\cite{vonBuddenbrock:2020ter,ATLAS:2023ajo,CMS:2023ftu}& Moderate $H_T$ & $t\overline{t}W^{\pm}, t\overline{t}t\overline{t}$ & $\approx 3\sigma$ \\ 
$\ell^\pm\ell^\pm, 3\ell, ($no$\;b$-jet)~\cite{Hernandez:2019geu,CMS:2021ixs,ATLAS:2022xnu} & In association with $h$ & $W^{\pm}h(125), WWW$ & $\gtrapprox4\sigma$ \\ 
$Z(\rightarrow\ell\ell)\ell, ($no$\; b$-jet)~\cite{Buddenbrock:2019tua,CMS:2022uhn} & $p_{\rm T}^Z<100$\,GeV & $ZW^{\pm}$ & $> 3\sigma$ \\ 
      \end{tabular}
      \end{center}
The fact that the leptons in these channels are non-resonant, i.e.~no peak in the invariant mass spectrum is observed, shows that, at least within the SM, they are related to leptonic $W$ decays. 

These excesses correspond to Higgs-like signatures (i.e.~$h\to WW$), which are experimentally robust and on the theoretical side higher-order QCD and EW corrections have been calculated for the main SM backgrounds. Most prominently, next-to-next-to-leading order QCD corrections are available for leptonic observables in $t\overline{t}$ distributions~\cite{Czakon:2020qbd}, non-resonant $W^+W^-$~\cite{Gehrmann:2014fva,Grazzini:2016ctr,Hamilton:2016bfu,Re:2018vac,Caola:2015rqy}, $ZW^{\pm}$~\cite{Grazzini:2017ckn}, $Wh$~\cite{Brein:2003wg,Ferrera:2011bk,Campbell:2016jau}, and $t\overline{t}W^{\pm}$~\cite{Buonocore:2023ljm} production. Electroweak corrections are also available at next-to-leading order and found to be small for Higgs-like signals~\cite{Ciccolini:2003jy,Denner:2011id,Denner:2016jyo,Biedermann:2016guo,Dittmaier:2019twg}. Furthermore, the description of the data by the SM outside these Higgs-like regions is within the residual errors. 

A particularly significant disagreement is observed in differential lepton distributions in $t\bar t$ measurements~\cite{Buddenbrock:2019tua,Banik:2023vxa} (see left diagram in Fig.~\ref{FeynmanDiagrams} f)). For all SM simulations used, ATLAS finds such a high $\chi^2$ value that they conclude~\cite{ATLAS:2023gsl}: ``No model (SM simulation) can describe all measured distributions within their uncertainties.'' While this effect warrants further investigation of the SM predictions, it is important to note that excesses also appear in $WW$ signatures without jets (where SM $t\overline{t}$ production is strongly suppressed) and in $Wh/3W$, $t\bar tW$, $t\overline{t}t\overline{t}$ and $ZW$ production with low $Z$-boson transverse momentum ($p^{Z}_{T}$), which indicates that the excess is likely not due to a mismodelling of the SM $t\overline{t}$ production and decay.\footnote{In addition, there is a hint for a resonant $t\bar t$ excess at around $400\,$GeV~\cite{CMS:2019pzc} with a local (global) significance of $3.5\sigma$ ($1.9\sigma$).}

The multi-lepton anomalies can be explained by the associated production of new scalars (i.e.~via the decay of a heavier scalar into two lighter ones). In particular, the deviations from the SM predictions in the differential distributions of leptons in the measurements of $t\bar t$ decays can be resolved by the production of a new neutral Higgs, $H$, that decays into two lighter ones $S$ and $S^\prime$ which subsequently decay to $W$ bosons and $b$ quarks, respectively~\cite{Banik:2023vxa} (see Fig.~\ref{FeynmanDiagrams} f)). This setup is preferred over the SM hypothesis by more than $5\sigma$ and points towards $m_S\approx150\,$GeV.
Similarly, the excess in $h\to WW$ can be described by a new Higgs boson decaying to $WW$~\cite{vonBuddenbrock:2017gvy,Coloretti:2023wng} and the same sign lepton signals with $b$ jets by the associated production of $H$ with top quarks where again $H\to SS^\prime\to WWbb$~\cite{Buddenbrock:2019tua}. 

Given the very large statistical significance of many channels of the multi-lepton anomalies (more than $8\sigma$ for the simplified model of Ref.~\cite{Buddenbrock:2019tua}) achieved with the run-2 data, the focus of studies with run-3 data (2022-2025) will shift to the signatures that are currently statistically limited. For instance, the study of the differential $\ell^+\ell^-$ distributions with a full jet veto or $\ell^\pm\ell^\pm$ with and without a $b$-jet would profit from more data. On the theory side, merging full next-to-next-to-leading order calculation~\cite{Czakon:2020qbd} and including off-shell effects~\cite{Jezo:2023rht} with parton showers at the same accuracy would significantly improve the SM simulations.

\subsection{Higgs-like resonant signals ($YY=\gamma\gamma,\tau\tau,WW,ZZ$)}

New particles that are directly produced at colliders show up as bumps in the otherwise continuous invariant mass spectrum of the corresponding decay products. For scalar bosons, di-photon distributions are very sensitive: Even though they have in general small rates because they are loop-suppressed, the experimental signature is very clear. In fact, there are several hints for di-photon resonances at $95\,$GeV~\cite{CMS:2023yay,ATLAS:2023jzc},
$\approx152$\,GeV~\cite{ATLAS:2021jbf}\footnote{The mass of this excess is consistent with invariant mass of di-leptons in the multi-lepton anomalies.} and also $\approx680\,$GeV~\cite{CMS:2017dib,ATLAS:2021uiz}. The hint at $95\,$GeV is supported by a di-taus excess reported by CMS~\cite{CMS:2022rbd} (however, not confirmed by ATLAS~\cite{ATLAS:2022yrq}), a $ZH$ signal (with $H\to b\bar b$) by LEP~\cite{LEPWorkingGroupforHiggsbosonsearches:2003ing} as well as the $WW$ channel~\cite{vonBuddenbrock:2017gvy,Coloretti:2023wng}. The $\gamma\gamma$ (plus missing energy) hint at 152\,GeV is supported by several signals in associated production~\cite{ATLAS:2023omk,ATLAS:2021jbf}, including $WW+$missing energy~\cite{Coloretti:2023wng}. Combining all channels, global significances of $3.8\sigma$ and $3.9\sigma$ are found for 95\,GeV~\cite{Bhattacharya:2023lmu} and 152\,GeV~\cite{Crivellin:2021ubm}, respectively, if for the latter, a simplified model with $pp\to H\to SS^*$ is assumed.\footnote{Ref.~\cite{Bhattacharya:2023lmu} updated the results of Ref.~\cite{Crivellin:2021ubm} by including additional new excesses, further increasing the significance of the narrow excess at around 152\,GeV. }

There are also hints for a new scalar in di-photon and di-$Z$ searches with a mass around 680$\,$GeV~\cite{ATLAS:2020tlo,ATLAS:2021uiz} (around $3\sigma$ each~\cite{Consoli:2022lyl}). Taking into account the resolution, this is compatible with the $3.8\sigma$ ($2.8\sigma$) local (global) excess in $bb\gamma\gamma$ at around 650\,GeV~\cite{CMS:2022tgk} (where the $b\bar b$ invariant mass is compatible with $95\,$GeV) and the $WW$ excess~\cite{CMS:2022bcb} in vector-boson fusion category.\footnote{Vector-boson fusion means that the new scalar is radiated from a $Z$ or $W$ pair. At the LHC, this leads to the presence of two forward jets which are thus close to the beam-line axis.} Furthermore. ATLAS seems an excess in $A\to H+Z$ with $H\to b\bar b,t\bar t$ at $m_A\approx 650$\,GeV and $m_H\approx450\,$GeV, with a local (global) significance of $2.85\sigma$ ($2.35\sigma$). However, the $bb\gamma\gamma$ is diminished by the non-observation of an excess in $\tau\tau\gamma\gamma$~\cite{CMS:2021yci} and the $WW$ excess cannot be fully explained within a model~\cite{LeYaouanc:2023zvi}.

These hints for resonances point towards the extension of the Higgs sector of the SM because only scalars can decay to photons. For the $95\,$GeV excess, at least an $SU(2)_L$ doublet~\cite{Haisch:2017gql}, triplet~\cite{Ashanujjaman:2023etj} or even a more complex scalar sector is needed~\cite{Biekotter:2022abc}. If one aims at also addressing the di-Higgs excess (i.e.~$650\,{\rm GeV}\to b\bar b(90\,{\rm GeV})+\gamma\gamma(125\,{\rm GeV})$)\footnote{The values in the brackets refer to the masses of the resonances which decay to the corresponding final states, i.e.~125\,GeV stands for the SM Higgs.}, the resonant pair production of the SM Higgs and a new scalar is required~\cite{Banik:2023ecr}. To account for the $152\,$GeV excess an even larger scalar sector is necessary since the bulk of the signal is in associated production. In fact, not only are the most significant excesses related to missing energy, the $WW$ signal can also be explained for $m_S\approx 150\,$GeV, i.e.~the decay chain $pp\to H\to (S\to \gamma\gamma,WW)+(S^\prime\to$invisible) describes data well. In general, a quite complicated scalar sector is suggested, like an extended Georgi-Machacek model~\cite{Georgi:1985nv,Kundu:2022bpy} or excitations of the SM Higgs~\cite{Consoli:2020kip}.

Given the current strength of the excesses, LHC Run~3, but at the latest the high-luminosity LHC~\cite{CidVidal:2018eel}, should suffice to verify or falsify the existence of these particles. However, to fully explore their properties, a $e^+e^-$ accelerator could be required. 

\begin{boldmath}
\subsection{(di-)di-jet resonances ($jj(-jj)$)}
\end{boldmath}

A particle decaying into two quarks (or two gluons) results in a di-jet event at the LHC. ATLAS~\cite{ATLAS:2018qto} observed a weaker limit than expected if there were no NP signal in resonant di-jet searches slightly below $1\,$TeV. Furthermore, CMS~\cite{CMS:2022usq} found hints for the (non-resonant) pair production of di-jet resonances with a mass of $\approx 950$\,GeV with a local (global) significance of 3.6$\sigma$ (2.5$\sigma$). This compatibility suggests that both excesses might be due to the same new particle $X$, once directly (resonantly) produced in proton-proton collisions ($pp\!\to\! X\!\to\! jj$), once pair produced via a new state $Y$ ($pp\!\to\! Y^{(*)}\!\to\! XX\!\to\! (jj)(jj)$). In fact, Ref.~\cite{Crivellin:2022nms} finds a global $3.2\sigma$ significance at $m_Y\approx 3.6\,$TeV. In the latest analysis, ATLAS finds a di-di-jet excesses~\cite{ATLAS:2023ssk} at $\approx3.3\,$TeV with a di-jet mass of $850\,$GeV which could be compatible with the CMS one once the quite poor jet energy resolution is taken into account. Furthermore, there is a slight excess in $tb$ searches at $\approx3.5\,$TeV~\cite{ATLAS:2023ibb}.

As explanations, two options come to mind~\cite{Crivellin:2022nms}: two scalar DQ (see Fig.~\ref{FeynmanDiagrams} g)) or new massive gluons seem to be the most plausible candidates~\cite{Crivellin:2022nms}. While the first one could explain the $tb$ excess, a specific realization of the latter is based on an $SU(3)_1\times SU(3)_2\times SU(3)_3$ gauge group, broken down to $SU(3)$ colour via two bi-triplets.
\medskip

\subsection{Non-resonant di-electrons ($q\bar q\to e^+e^-$)}

If the mass of a particle exceeds the energy reach of a collider, its impact can still be seen by looking at the high-energetic end of the spectrum of a distribution where such effects are most relevant because they possess a relative enhancement w.r.t.~the SM (see left-diagram in Fig.~\ref{FeynmanDiagrams} h)). In such a non-resonant search for high-energetic oppositely charged leptons, CMS and ATLAS observe more electrons than expected in the SM~\cite{CMS:2021ctt,ATLAS:2020yat}.\footnote{Similarly significant excesses have been observed in the corresponding resonant searches. However, here the look-elsewhere effect reduces the significance which is not the case for non-resonant searches.} Because the number of observed muons is compatible with the SM prediction, this is a sign of lepton flavour universality violation and the ratio of muons over electrons provided by CMS has the advantage of reduced theoretical uncertainties~\cite{Greljo:2017vvb}. Performing a model-independent fit, one finds the NP at a scale of $10\,$TeV with order one couplings can improve over the SM hypothesis by $\approx3\sigma$~\cite{Crivellin:2021rbf}.

As this analysis involves non-resonant electrons that do not originate from the on-shell production of a new particle, NP must be heavier than the energy scale of the LHC (or be produced non-resonantly like LQs). This can be achieved with NP at the 10\,TeV scale with order one coupling to first-generation quarks and electrons~\cite{Crivellin:2022rhw}. Therefore, $Z^\prime$~bosons~\cite{Crivellin:2021bkd} or LQs~\cite{Crivellin:2021egp} (see Fig.~\ref{FeynmanDiagrams} h)) have the potential to explain the CMS measurement. LHC Run~3 should suffice to determine the validity of these excesses.

\begin{figure}
	\centering
	\vspace{-3mm}
	\includegraphics[width=0.8\textwidth]{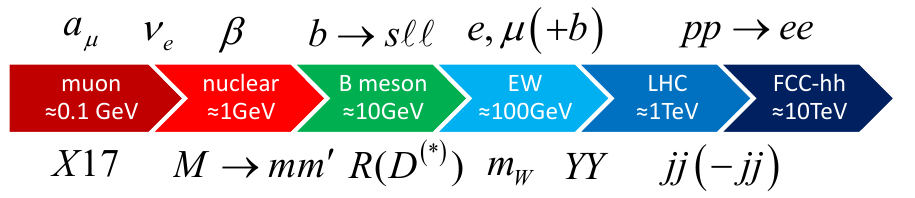}
	\vspace{-3mm}
	\caption{Compilation of various anomalies ordered according to the corresponding energy scale. }
	\label{anomalies}
\end{figure}

\begin{figure}[t]
    \centering    
    \includegraphics[scale=0.8]{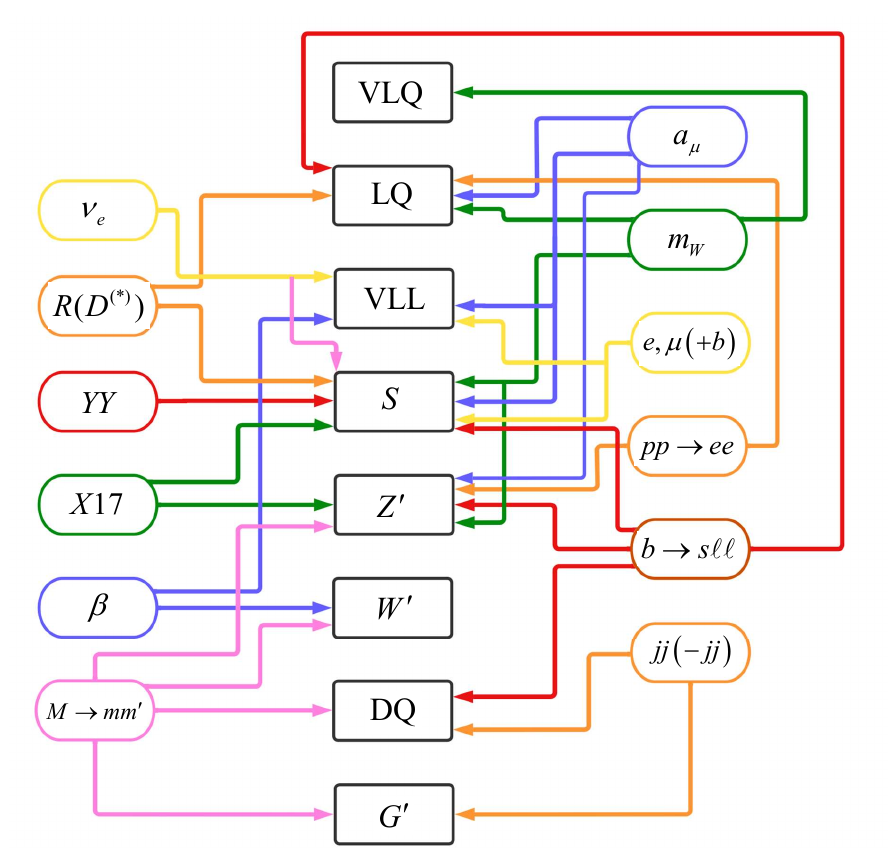}
    \caption{Summary of the anomalies together with the implications for extending the SM with new particles: Leptoquarks (LQ), vector-like leptons (VLL), vector-like quarks (VLQ) electrically neutral scalars ($S$), neural gauge bosons ($Z^\prime$), charged gauge bosons ($W^\prime$) and di-quarks (DQ).}
    \label{summary}
\end{figure}

\section{Comparison, conclusions and outlook}

The anomalies observed in particle physics are summarized in Fig.~\ref{anomalies}, together with their corresponding energy scale, showing that they range over at least five orders of magnitude. While one cannot expect that all anomalies will be confirmed, it is also statistically unlikely that all will disappear. Therefore, it is important to investigate their implications for NP in order to assess possible correlations among them and identify signatures for future verification (or falsification). 

Let us now compare the anomalies concerning their experimental and theoretical features:\footnote{Please note that even though we try to be objective here, the impact of personal opinion is unavoidable.}
\begin{itemize}
  \item Anomalous magnetic moment of the muon ($a_\mu$):
    \newline    \textcolor{green}{\bf +} Precise and confirmed direct measurements 
    \newline    \textcolor{red}{\bf --} Standard Model prediction plagued by hadronic uncertainties
    \newline    \textcolor{red}{\bf --} Tensions within $e^+e^-\to$hadrons measurements and with lattice QCD
    \newline    \textcolor{red}{\bf --} Quite large NP effect needed; model building is challenging
\item The 17\,MeV Anomaly in excited nuclei decays ($X17$)    \newline    \textcolor{green}{\bf +} High statistical significance
    \newline\textcolor{red}{\bf --} Only observed by a single experiment (despite different setting)
    \newline\textcolor{red}{\bf --} Possibility of not understood nuclear effects
\item Anomalies in electron appearance and disappearance ($\nu_e$)
         \newline\textcolor{green}{\bf +} High statistical significance
             \newline\textcolor{green}{\bf +} Observed by different experiments
            \newline\textcolor{red}{\bf --} Theory errors might be underestimated
           \newline\textcolor{red}{\bf --} Explanation via sterile neutrino mixing excluded by other experiments
  \item Beta-decay Anomalies ($\beta$):
      \newline  \textcolor{green}{\bf +} Natural place to search for NP since only sub-permille effect w.r.t.~the SM is needed 
    \newline    \textcolor{red}{\bf --} Only one (competitive) measurement of $K\to\mu\nu$ available
     \newline   
    \textcolor{red}{\bf --} Beta decays need hadronic theory input to extract $V_{ud}$
            \newline \textcolor{red}{\bf --} Lifetime difference can only be explained by exotic NP
  \item Hadronic meson decays ($M\to mm^\prime$):
        \newline  \textcolor{green}{\bf +} Many different channels
                   \newline \textcolor{red}{\bf --} SM plagued by hadronic effects
                \newline \textcolor{red}{\bf --} NP explanations challenged by LHC searches
    \item Flavour changing neutral current $B$ decays ($b\to s\ell^+\ell^-$)
    \newline  \textcolor{green}{\bf +} Many different observables measured
    \newline  \textcolor{green}{\bf +} Consistent picture
    \newline  \textcolor{green}{\bf +} Large significance
    \newline  \textcolor{green}{\bf +} Possible connection to $b\to c\tau\nu$
    \newline    \textcolor{red}{\bf --} Sensitive to form factors and other hadronic input
        \item Charged current tauonic $B$ decays ($R(D^{(*)})$)
    \newline  \textcolor{green}{\bf +} Measurements from different collaborations
    \newline  \textcolor{green}{\bf +} Small theory uncertainty
    \newline  \textcolor{green}{\bf +} Possible connection to $b\to s\ell^+\ell^-$
    \newline    \textcolor{red}{\bf --} Difficult measurement
    \newline    \textcolor{red}{\bf --} Limited significance
    \newline    \textcolor{red}{\bf --} Large effect needed; challenging model building
        \item $W$ mass ($m_W$)
    \newline  \textcolor{green}{\bf +} Theoretically clean
    \newline  \textcolor{green}{\bf +} Statistically significant
    \newline  \textcolor{green}{\bf +} Very sensitive to NP; many natural NP explanations
    \newline    \textcolor{red}{\bf --} Tensions among the measurements
        \item LHC Multi-lepton anomalies ($e\mu(+b)$):
    \newline  \textcolor{green}{\bf +} Statistically very significant
    \newline  \textcolor{green}{\bf +} Large multiplicity of signatures
    \newline  \textcolor{green}{\bf +} Coherent picture
    \newline  \textcolor{green}{\bf +} Consistent with the Higgs-like signals
    \newline    \textcolor{red}{\bf --} Some of the SM predictions can be difficult
    \newline    \textcolor{red}{\bf --} Complex SM extension needed 
            \item Higgs-like signal ($YY$)
    \newline  \textcolor{green}{\bf +} Statistically significant
    \newline  \textcolor{green}{\bf +} Many different channels
    \newline  \textcolor{green}{\bf +} Motivated by the multi-lepton anomalies
    \newline    \textcolor{red}{\bf --} Possible look-elsewhere effect
            \item (di-)di-jet ($jj(-jj)$)
    \newline  \textcolor{green}{\bf +} Agreement between different measurements
    \newline    \textcolor{red}{\bf --} Poor mass resolution
    \newline    \textcolor{red}{\bf --} Challenging theory explanation
        \item Non-resonant di-electrons ($q\bar q\to e^+e^-$):
    \newline  \textcolor{green}{\bf +} Agreement between ATLAS and CMS
    \newline  \textcolor{green}{\bf +} Ratio theoretically clean
    \newline    \textcolor{red}{\bf --} Limited statistics
    \newline    \textcolor{red}{\bf --} Electrons are difficult LHC signatures
\end{itemize}

The anomalies are also compared in Table~\ref{tab:summary} w.r.t.~several criteria which try to answer the following questions:
\begin{itemize}
    \item Experimental signature: Is the experimental environment clean? Is the signal well separated from the background?
    \item Experimental consistency: Do multiple independent measurements exist? Are they in agreement with each other?
        \item Standard Model prediction: How accurate and reliable is the SM prediction? Are the conflicting results?
    \item Statistical significance: How sizable are the deviations from the SM predictions?
    \item NP explanation: Are there models that can naturally account for the excess? Are they in conflict with other observables?
    \item Consistent connection: Are there connections to other anomalies via the same new particle or model? How direct is this connection?
\end{itemize} 
Here, \textcolor{green}{\bf +} reflects a positive assessment, \textcolor{red}{\bf --} a negative one and \textcolor{gray}{\bf 0} means neutral, i.e.~positive and negative aspects compensate to a good approximation.

\begin{table}
    \centering
    \begin{tabular}{c | c c c c c c}
& 
$\begin{array}{c}
{\rm{Exp}}{\rm{.}}\\
{\rm{signature}}
\end{array}$
&
$\begin{array}{c}
{\rm{Exp}}{\rm{.}}\\
{\rm{consistency}}
\end{array}$
&
$\begin{array}{c}
{\rm{SM}}\\
{\rm{prediction}}
\end{array}$
&
$\begin{array}{c}
{\rm{statistical}}\\
{\rm{significance}}
\end{array}$
&
$\begin{array}{c}
{\rm{NP}}\\
{\rm{explanation}}
\end{array}$&$\begin{array}{c}
{\rm{consistent}}\\
{\rm{connection}}
\end{array}$
\\
\hline
{{$a_\mu$ }} &  \color{green}{\bf +} &  $\;\;\;$\color{gray}{\bf 0}\footnote{Note that while the results of the Brookhaven and Fermilab experiments for $a_\mu$ agree very well, the neutral rating it due to the inconsistencies between the experiments measuring $e^+e^-\to$hadrons which are used for calculating the SM prediction via dispersion relations.} &  \color{red}{\bf --} &  \color{green}{\bf +} &  \color{gray}{\bf 0} &  \color{red}{\bf --} \\
{{$X17$}} &   \color{green}{\bf +} & \color{gray}{\bf 0}& \color{red}{\bf --}&  \color{green}{\bf +} &  \color{gray}{\bf 0} &  \color{gray}{\bf 0} \\
{{$\nu_e$}} &   \color{red}{\bf -} & \color{gray}{\bf 0}& \color{red}{\bf --}&  \color{green}{\bf +} &  \color{red}{\bf --} &  \color{red}{\bf --} \\
{{$\beta$}} &   \color{green}{\bf +} & \color{gray}{\bf 0}& \color{gray}{\bf 0}&  \color{red}{\bf --} &  $\;\;\;\;$\color{green}{\bf +} (\color{red}{\bf -})\footnote{The first assessment refers to the deficit in CKM unitarity and the $V_{us}$ disagreement while the second one in brackets refers to the lifetime difference. } &  \color{green}{\bf +} \\

{{$M\to mm^\prime$}} &   \color{gray}{\bf 0} & \color{green}{\bf +}& \color{red}{\bf --}&  \color{gray}{\bf 0} &  \color{red}{\bf --} &  \color{gray}{\bf 0} \\

{$b \to s\ell \ell$ } &   \color{green}{\bf +} &  \color{green}{\bf +} & \color{gray}{\bf 0}&  \color{green}{\bf +} & \color{gray}{\bf 0}&  \color{green}{\bf +} \\
{$R(D^{(*)})$ } &   \color{red}{\bf --} &  \color{green}{\bf +} &  \color{green}{\bf +} & \color{red}{\bf --}&  \color{red}{\bf --} &  \color{green}{\bf +} \\
{{$m_W$}} &  \color{gray}{\bf 0}&  \color{red}{\bf --} &  \color{green}{\bf +} &  \color{green}{\bf +} &  \color{green}{\bf +} &  \color{green}{\bf +} \\
{$e\mu \left( {  + b} \right)$} & \color{gray}{\bf 0}&  \color{green}{\bf +} & \color{gray}{\bf 0}&  \color{green}{\bf +} &\color{gray}{\bf 0}&  \color{green}{\bf +} \\
{$YY$} &   \color{green}{\bf +} &  \color{green}{\bf +} &  \color{green}{\bf +} & \color{gray}{\bf  \color{gray}{\bf 0}}&  \color{green}{\bf +} &  \color{green}{\bf +} \\
{$jj( {-jj} )$} &  \color{gray}{\bf 0}&  \color{green}{\bf +} &  \color{green}{\bf +} & \color{gray}{\bf 0}& \color{gray}{\bf 0}&  \color{red}{\bf --} \\
{$pp \to ee$} &  \color{gray}{\bf 0}&  \color{green}{\bf +} &  \color{green}{\bf +} &  \color{red}{\bf --} & \color{gray}{\bf 0}&  \color{red}{\bf --} 
\end{tabular}
    \caption{Comparison of the different anomalies in particle physics in terms of various features. See main text for details. }
    \label{tab:summary}
\end{table}

The anomalies discussed above, together with the extensions of the SM to which they point, are shown in Fig.~\ref{summary}. One can see that many extensions point towards new Higgs-like scalars. In particular, the agreement between the mass of the scalar suggested by the multi-lepton anomalies and the $\gamma\gamma$ excess around $152\,$GeV is striking. LQ are also interesting candidates and, in particular, allow for a combined and correlated explanation of $b\to c\tau\nu$ and $b\to s\ell^+\ell^-$ via the tau-loop~\cite{FernandezNavarro:2022gst,Aebischer:2022oqe}. Finally, $Z\to b\bar b$, $m_W$ and the CAA could be explained by VLQ. Of course, in a UV complete model, many more possible connections exist, as can be seen from Fig.~\ref{summary}. This offers interesting open research directions. 

Particle physics is currently a very exciting area of research. While the SM has been consolidated over the last five decades, hints of new particles and new interactions are emerging. Despite originating from very different experiments and ranging over five orders of magnitude in energy, the task is to find combined explanations to verify or falsify the predictions of the anomalies in the future. However, one has to take into account that most likely not all anomalies will be confirmed by ongoing and forthcoming experimental efforts. Nonetheless, already establishing one of these hints beyond a reasonable doubt would lead particle physics into a new era, the BSM age. 

\bibliography{mainv2}

\section*{Acknowledgements}
The work of A.C.~is supported by a professorship grant from the Swiss National Science Foundation (No.\ PP00P2\_211002). B.M.~gratefully acknowledges the South African Department of Science and Innovation through the SA-CERN program, the National Research Foundation, and the Research Office of the University of the Witwatersrand for various forms of support. The authors want to thank William Murray for pointing to excess in the $Zhh$ final state and Lindsay Donaldson for assistance with the prospects of investigating the $X17$ anomaly.

\section*{Author contributions}
The main part of the writing was done by A.C.~while B.M.~contributed to the part concerning the definition of anomalies, multi-lepton anomalies, the $X17$ excess, the di-di-jet excesses and the Higgs-like signals.



\end{document}